\documentstyle[12pt]{article}
\newcommand{\lraw}{\leftrightarrow}
\newcommand{\be}{\begin{equation}}
\newcommand{\ee}{\end{equation}}
\newcommand{\ba}{\begin{eqnarray}}
\newcommand{\ea}{\end{eqnarray}}
\newcommand{\bann}{\begin{eqnarray*}}
\newcommand{\eann}{\end{eqnarray*}}

\renewcommand{\hfill}{\hspace*{\fill}}
\newcommand{\bepsilon}{\mbox{\boldmath $\epsilon$}}
\newcommand{\bell}{\mbox{\boldmath $\ell$}}
\newcommand{\bold}[1]{\mbox{\boldmath $#1$}}
\begin{document}
\hbadness=10000
\thispagestyle{empty}
\begin{samepage}
\title{\hfill {\normalsize DMR-THEP-93-2/W}\\*[-1.5ex]
\hfill {\normalsize hep-ph/9308245 }\\*[-1.5ex]
\hfill {\normalsize May 1993} \\[2ex]
Dynamical properties of heavy-ion collisions from the
Photon-Photon intensity correlations}
\author{L.V. Razumov\thanks{E.Mail:
razumov@convex.hrz.Uni-Marburg.de},\ \ \
R.M. Weiner\thanks{E. Mail: WEINER@vax.hrz.Uni-Marburg.de}}

\date{Physics Department, University of Marburg, Marburg, F.R. Germany}
\maketitle
\thispagestyle{empty}
\begin{center}
{\large\bf Abstract}\\[2ex]
\end{center}
\noindent
We consider here the bremsstrahlung emission of photons at low and
intermediate energies $ E_{lab}\le 1000 MeV/u$ of the projectile.  and
derive expressions more general than previous results obtained by
Neuhauser which were limited to the case of isotropic systems. We find
that the two-photon correlation function strongly depends not only on
the space-time properties of the collision region but also on the
dynamics of the proton-neutron scattering process in nuclear matter.
As a consequence of polarisation correlations it turns out that for a
purely chaotic source the intercept of the correlation function of
photons can reach the value $3$ (as compared with the maximum value
$2$ for isotropic systems).  Furthermore even for ``hard" photons
($E_{\gamma} > 25 MeV$) the maximum of the correlation function can
reach the value of $2$ in contrast with the value of $1.5$ derived by
Neuhauser for this case.  The formulae obtained in this paper which
include also the possible presence of a coherent component can be used
as a basis for a systematic analysis of photon
intensity-interferometry experiments.
\end{samepage}
\newpage
\section{Introduction}

The Hanbury-Brown and Twiss intensity interferometry \cite{HBT} plays
at present an important role in particle and nuclear physics in a wide
range of c.m. energies, because this method can provide information
about the space-time properties of the sources.  From hadron
interferometry one obtains information mainly about the properties of
the fireball just near the freeze-out. Moreover, in this case we are
faced with some dynamical problems like the final state interaction
and contribution from resonances which mask the real space-time size
of the collision region. In contrast, photons and leptons have a large
mean free path and leave the collision region just after their
production. Additionally we can assume that photons have no final
state interaction (light-light scattering can be neglected).
Therefore direct photons (i.e. photons not originating from $\pi^0$
decay) and dileptons are good probes for the earliest stages of the
reaction.  Another advantage of electro-magnetic probes is that their
emission is controlled by QED rather than by strong interaction
dynamics for which no comparable theory exists yet.

In this paper we consider mostly the emission of ``hard'' photons
(their definition will be given below) in heavy ion reactions in the
energy range up to 1 Gev/u. We will show that the photon-photon
correlation function can provide there not only space-time information
but also some dynamical information about the proton-neutron
scattering in nuclear matter. It is known
\cite{GSI,C}
that at intermediate energies hard photons are produced mainly through
$bremsstrahlung$ and $\pi_0$ decay. Since the photon pairs from the
$\pi_0$ decay can be eliminated to a great extent by measuring the
photon pair invariant mass, we concentrate ourselves on
$bremsstrahlung$ photons.
\section{Bremsstrahlung photons and their correlations}

The photon spectrum is usually described by a superposition of two
contributions, both parametrized by exponentials \cite{GSI,C} and
referring to different photon energies respectively: the low energy
part is assumed to be due to thermal photons and the slope is just
given by the temperature. The second (high energy) part can be
considered \cite{GSI} as due to a mixture of bremsstrahlung (see
Chapter \ref{ef}) and photons originating from decays (mostly $\pi_0
\to \gamma\gamma$). Measurements of the angular distribution of
radiation, the photon-source velocity
\cite{GSI,C}
as well as the impact-parameter dependence of the photon multiplicity
\cite{GANIL.c} suggest that the direct hard photons mainly originate
from bremsstrahlung in independent proton-neutron
collisions\footnote{The photon emission from proton-proton collisions
is of quadrupole form and therefore highly suppressed as compared with
the dipole radiation from proton-neutron collisions.}.

As was shown in \cite{C} the non-relativistic recoilless
bremsstrahlung formula for the current in a\ $p-n$\
collision\footnote{The experimently found one-photon spectra are of
exponential form \cite{GSI,C}. This property can be also incorporated
into this model, see Chapter \ref{ef}.}
\be j^{\lambda}(k) =
\frac{ie}{mk^0}{\bf p}\cdot \mbox{\boldmath $\epsilon$}_{\lambda}(k)
\label{eq:j}
\ee
(where ${\bf p}={\bf p}_i-{\bf p}_f$ is the difference
between the initial and final momentum of the proton,
$\bepsilon_{\lambda}(k)$ is the vector of linear polarisation and $k$
the 4-momentum) works well even in the relativistic case due to the
fact that relativistic and recoil corrections to some extent
compensate each other.  If we have $N$ such sources we can write down
the transition current as follows:
\be
J^{\lambda}(k) = \sum^N_{n=1} e^{iky_n} j^{\lambda}_n(k)
\label{eq:jj}
\ee
The index $n$ here labels the independent $p-n$ collisions taking
place at different space-time points $y_n$.  Formulas (\ref{eq:j}),
(\ref{eq:jj}) are examples of radiation by classical currents, where
the influence of the emitted photons on the $p-n$ collision process is
negligible.

For an arbitrary classical current\ $J^\mu(x)$\ coupling with the
photon field
\ba
&& A^\mu(x) \ = \ \int \tilde{dk} \left(a_\lambda(k)\epsilon^\mu_\lambda(k)
{\rm e}^{-ikx} \ + \ a^{^{+}}_\lambda(k)\epsilon^{^{\!\! *}\mu}_\lambda(k)
{\rm e}^{ikx}\right) \quad ,   \\
&& \tilde{dk} \ \equiv \ d^3k/[(2\pi)^32k_0] \  \ , \nonumber \\
&& \left[ a_{\lambda_1}(k_1)\ ,\ a^{^{+}}_{\lambda_2}(k_2)
\right] \ \equiv \
\delta_{\lambda_1\lambda_2}(2\pi)^3\cdot 2k_0\delta^3({\bf k}_1-{\bf k}_2)
\nonumber
\label{eq:A}
\ea
through the conventional interaction lagrangian
\be
L_{int} \ = \ -J_\mu(x)A^\mu(x) \label{eq:L}
\ee
we can easily find the exact\  $S$-matrix
\be
S =  {\rm exp}\left[ -\frac{1}{2}\sum^2_{\lambda=1}\int \tilde{dk}
\left| J^\lambda(k) \right|^2 \right]:{\rm exp}\left[ -i \int
\tilde{dk} \left(J^\lambda(k)a^{^{+}}_\lambda(k)+ J^{^{+} \lambda}(k)
a_\lambda(k)\right) \right] :
\label{eq:SM}
\ee
which contains all the information on the photon production and absorption.
We use in (\ref{eq:SM}) the notation:
\be
J^\lambda(k) \ \equiv \ \epsilon^{^{\!\!
*}\lambda}_\mu(k)\tilde{J}^\mu(k) \ \ .
\label{eq:JL}
\ee
where $\tilde{J}^{\mu}(k)$ is the Fourier transform of the
current $J^{\mu}(x)$. Before the collision there are no photons,
i.e. we have the photon vacuum in the initial state. \\
The amplitude to produce\ $n$-photons from the photon vacuum by  the
classical current $J$ follows directly from  (\ref{eq:SM}):
\be
<0|\prod^n_{i=1}a_{\lambda_i}(k_i)S|0>\ =\ <0|S|0>\prod^n_{i=1}
\left(-iJ^{\lambda_i}(k_i)\right)\ \ .
\label{eq:AMP}
\ee
The corresponding cross-section can be written in a condensed and
convenient form by means of the generating functional for the
exclusive processes. Let us introduce the notation $q\equiv ({\bf
k},\lambda)$\ for the momentum \ $k$\ and polarisation
\ $\lambda$\  degrees of freedom of photon.
Integration over \ $q$\ means integration over momentum (with the
invariant measure \ $\tilde{dk}$\ ) and summation over polarisation \
$\lambda$\ . We define the exclusive generating functional \
$g^{ex}[Z]$\ as follows:

\be
g^{ex}[Z]\ \equiv \ \sum^\infty_{n=0}\frac{1}{n!}\int\prod^n_{i=1}dq_i
Z(q_i)\frac{1}{\sigma^{ex}}\cdot\frac{d^n\sigma^{ex}}{dq_1\times\ldots
\times dq_n}
\label{eq:gex}
\ee
The conservation of probability gives the normalisation\
$g^{ex}[Z=1]=1$\ .  The method of the generating functionals allows
one to avoid the complicated combinatorics which usually appears when
one consider inclusive processes. The generating functional for the
inclusive cross-section can be introduced in a similar way
\be
g^{in}[Z]\ \equiv \ \sum^\infty_{n=0}\frac{1}{n!}\int\prod^n_{i=1}dq_i
Z(q_i)\frac{1}{\sigma^{in}}\cdot\frac{d^n\sigma^{in}}{dq_1\times\ldots
\times dq_n}
\label{eq:gin}
\ee
(with the normalisation condition \ $g^{in}[Z=0]=1$\ ) and can be very
simple expressed through the exclusive one
\be
g^{in}[Z]\ =\ g^{ex}[Z+1] \ \ .
\label{ginex}
\ee
In our case (see eq.(\ref{eq:AMP})) the summation over \ $n$\ can be
performed explicitly and we are left with the simple expressions
\ba
g^{ex}[Z]&=& {\rm
exp}\left[\sum^2_{\lambda=1}\int\tilde{dk}|J^\lambda(k)|^2
\left(Z_\lambda(k)-1\right) \right]  \\
\label{eq:gexs}
g^{in}[Z]&=& g^{ex}[Z+1]\ =\ {\rm
exp}\left[\sum^2_{\lambda=1}\int\tilde{dk}
|J^\lambda(k)|^2Z_\lambda(k) \right] \label{eq:gins}
\ea
If the classical current obeys some random behaviour which is the
present case, as we consider chaotic sources, the generating
functional should be subjected to the averaging over the current
distribution
\be
G^{in}\ =\ \left< g^{in}[Z] \right> \ \ .
\label{Gin:df}
\ee
Taking the variational derivatives of \ $G^{in}[Z]$\ with respect to \
$Z$\ one gets the inclusive spectra. For example single and double
inclusive spectra read as follows:
\ba
\rho^\lambda_1(k)&=& \left. \frac{\delta G^{in}[Z]}
{\delta Z_\lambda(k)}\right|_{Z=0} \ =\ \left\langle |J^\lambda(k)|^2
\right\rangle \label{rho1:df}  \\
\rho^{\lambda_1\lambda_2}_2(k)&=& \left.\frac{\delta^2 G^{in}[Z]}
{\delta Z_{\lambda_1(k_1)} \delta Z_{\lambda_2(k_2)}}\right|_{Z=0} \ =
\left\langle|J^{\lambda_1}(k_1)J^{\lambda_2}(k_2)|^2   \right\rangle
\label{rho2:df}
\ea
An essential ingredient in eqs. (\ref{Gin:df}) - (\ref{rho2:df}) is
the average prescription \ $<...>$.

For the model considered here, where hard bremsstrahlung photons
originate mostly from independent proton-neutron collisions at the
early stage of the heavy-ion reaction, the averaging prescription \
$<...>$\ appears quite naturally and consists of two parts:
\begin{enumerate}
\item The points  \ $y_n$\ where the independent  \ $p-n$\ collisions take
place (see eq. (\ref{eq:jj})) are considered to be randomly
distributed in the space-time volume of the source (e.g. fireball)
with a distribution function \ $f(y)$\ for each \ $p-n$\ collision.
This kind of averaging is typical for Bose-Einstein correlation
studies.
\item The amplitude of photon production  (\ref{eq:j}) is very sensitive
to the momentum transfer of the proton in the proton-neutron
collision.  That is why we have to take into account also the
fluctuations of the initial momentum of the proton: \ ${\bf p}_i={\bf
p}_0 + \Delta {\bf p}_F$\ .  Here ${\bf p}_0$ is the momentum of the
nucleus as a whole and $\Delta {\bf p}_F$ is the Fermi-motion of a
nucleon. The final momentum distribution of the proton \ $({\bf
p}_f)$\ can be in principle determined by dynamical models of
nucleus-nucleus collisions, e.g. BUU \cite{C}, etc.
\end{enumerate}
Since in practice we need only the first two moments of this
distribution we keep them as a free phenomenological parameters to be
determined in the experiment. The comparison of the experimental
values with model predictions can serve as an indirect test of the
model.

The fundamental quantity in our approach is the two-current correlator
in momentum space:
\begin{samepage}
\ba
&& <J^{\lambda_1}(k_1) J^{^{*} \lambda_2}(k_2)> = <J^{\lambda_1}(k_1)
J^{\lambda_2}(-k_2)> = \sum^N_{n,m=1} \!
\int \! \prod^N_{l=1} d^4 y_l f(y_l) \times \nonumber \\
&& {\rm exp}(ik_1y_n-ik_2y_m)
<j^{\lambda_1}_{n}(k_1)j^{\lambda_2}_{m}(-k_2)> = \!\! \sum^N_{n=1}
\left[ \tilde{f}(k_1-k_2) <j^{\lambda_1}_{n}(k_1)
j^{\lambda_2}_{n}(-k_2)> \right.  \nonumber \\ && - \left.
\tilde{f}(k_1) \tilde{f}(-k_2) <j^{\lambda_1}_n(k_1)>
<j^{\lambda_2}_{n}(-k_2)> \right] + <J^{\lambda_1}(k_1)>
<J^{\lambda_2}(-k_2)>\ \ .
\label{eq:F}
\ea
\end{samepage}
Here we use the properties $(J^{\lambda}(k))^{^{\! *}}=
J^{\lambda}(-k)$, \
$<j^{\lambda_1}_n(k_1)j^{\lambda_2}_m(-k_2)>=$\mbox{$<j^{\lambda_1}_n(k_1)>
<j^{\lambda_2}_m(-k_2)>$}\ for \ $n \ne m$\ (that is the hypothesis of
independent proton-neutron collisions), and \ $\tilde{f}(k)$\ denotes
the Fourier transform of \ $f(y)$\ with the normalisation \
\mbox{$\tilde{f}(k=0)=1\ $}. \\ Usually the function \ $\tilde{f}(k)$\ has a
steep maximum around \ $k=0$\ with a width of the order of the size of
the source \ $R\ $.  In the region of the Bose-Einstein peak for hard
photons in particular for $2kR\gg1$ we can therefore neglect $\
\tilde{f}(k_1)
\tilde{f}(-k_2) \ $ as compared to $\ \tilde{f}(k_1-k_2)\ $. This
approximation strongly simplifies the algebra without influencing
significantly the accuracy of the calculations.

\section{Chaotic sources}
Let us consider the case of a chaotic source for which $\
<J^{\lambda}_{ch}(k)>=0\ $.  The current correlator (\ref{eq:F}) is
now given by the simple formula:
\ba
\label{eq:F2}
<J^{\lambda_1}(k_1) J^{\lambda_2}(-k_2)> &=&
F^{\lambda_1\lambda_2}(k_1,k_2) \equiv \tilde{f}(k_1-k_2) \sum^N_{n=1}
<j^{\lambda_1}_{n}(k_1)j^{\lambda_2}_{n}(-k_2)> \nonumber \\ &=&
\tilde{f}(k_1-k_2)\frac{e^2/m^2}{k^0_1k^0_2}
\bepsilon^i_{\lambda_1}(k_1)\left( \sum^N_{n=1}<{\bf p}^i_n {\bf
p}^j_n>
\right) \bepsilon^j_{\lambda_2}(k_2) \quad .
\ea
$<{\bf p}^i_n {\bf p}^j_n>$\ denotes here the averaging with respect
to the distribution of the momentum transfer of the proton in the
collision $\ n \ $ and $\ \sum^{N}_{n=1}\ $ goes through all the
relevant proton-neutron collisions. As mentioned above the quantity $\
<{\bf p}^i_n {\bf p}^j_n>\ $ can be extracted from dynamical models of
heavy-ion collisions. But one can reach important and general
conclusions without specifying this quantity as follows.  We use the
axial symmetry around the beam direction.  The tensor decomposition of
$\ <{\bf p}^i_n {\bf p}^j_n>\ $ gives then:
\be
<{\bf p}^i_n {\bf p}^j_n>\ =\ \frac{\sigma_n}{3}\delta^{ij}\ +\
\delta_n \bell^i \bell^j \quad ,
\label{pp}
\ee
where $\ \bell\ $ is the unit vector in the beam direction and $\
\sigma_n\ $, $\ \delta_n\ $ are real positive constants.  Note that
this expression is more general than the corresponding one used in
\cite{N} where because of the isotropy assumption $\delta_n$ was assumed to
vanish. This generalization has important consequences to be exhibited
below.  Now let us separate the average value of the proton momentum
\mbox{transfer $\ <{\bf p}_n>\ $:}
\be
{\bf p}_n\ =\ \Delta{\bf p}_n\ +\ <{\bf p}_n>
\label{p}
\ee
We have then:
\be
<{\bf p}^i_n {\bf p}^j_n>=<\Delta {\bf p}^i_n\Delta {\bf p}^j_n>+
<{\bf p}^i_n><{\bf p}^j_n>= <\Delta {\bf p}^i_n\Delta {\bf p}^j_n> +
<{\bf p}>^2 {\bf \bell}^i{\bf \bell}^j
\label{pp:2}
\ee
where the tensor $\ <\Delta {\bf p}^i_n\Delta {\bf p}^j_n>\ $ can be
represented (due to axial symmetry) again as the sum of the two terms:
\be
<\Delta {\bf p}^i_n \Delta {\bf p}^j_n>\ =\ \frac{\sigma_n}{3}
\delta^{ij}\ +\
\xi_n \bell^i \bell^j \quad . \label{Dpp}
\ee
In order to find the coefficients $\ \sigma_n,\ \xi_n,\ \delta_n \ $
we split the 3-vectors in the transverse and the longitudinal parts:
\be
\Delta{\bf p}^{l}_n\ =\ {\bf \bell}\cdot({\bf \bell}\cdot\Delta{\bf p}_n);
\qquad
\Delta{\bf p}^{t}_n\ =\ \Delta{\bf p}_n\ -\ \Delta{\bf p}^{l}_n \qquad .
\label{p:lt}
\ee
The coefficients in (\ref{pp}), (\ref{Dpp}) read :
\ba
\frac{\sigma_n}{3}\ &=&\ \frac{1}{2}<(\Delta{\bf p}^{t}_n)^2>\ ; \qquad
\xi_n\ =\  <(\Delta{\bf p}^{l}_n)^2> - \frac{\sigma_n}{3}\ ; \nonumber \\
\delta_n\ &=&\ <(\Delta{\bf p}^{l}_n)^2> - \frac{\sigma_n}{3}+<{\bf p}_n>^2
\ \ . \label{parn:df}
\ea
With the help of (\ref{pp}), (\ref{parn:df}) one can express the
current correlator (\ref{eq:F2})
\be
F^{\lambda_1\lambda_2}(k_1,k_2)=\tilde{f}(k_1-k_2) \frac{N \cdot
e^2/m^2} {k^0_1k^0_2}\bepsilon^i_{\lambda_1}(k_1)
\left[\frac{\sigma}{3} \delta^{ij} +
\delta \bell^i \bell^j \right] \bepsilon^j_{\lambda_2}(k_2) \ ;
\label{eq:F3}
\ee
through two parameters $\ \sigma\ $ and $\ \delta\ $
\ba
\frac{\sigma}{3}\ &=&\  \frac{1}{N}\sum^N_{n=1}\frac{1}{2}
<(\Delta{\bf p}^{t}_n)^2> \ \; \\
\delta \ &=&\ \frac{1}{N}\sum^N_{n=1}[<(\Delta{\bf p}^{l}_n)^2> +
<{\bf p}_n>^2] - \frac{\sigma}{3} \ \ ,
\label{par:df}
\ea
which absorb the relevant dynamical information about the
proton-neutron scattering in the medium. The parameters $\ \sigma\ $
and $\ \delta\ $ can be extracted from experimental data on
photon-photon Bose-Einstein Correlations (cf. below) and/or calculated
from dynamical models for heavy-ion collisions.

The single inclusive cross-section for a detector, which does not
measure polarizations follows directly from (\ref{rho1:df}),
(\ref{eq:F3})
\be
\rho_1(k) =\sum^2_{\lambda=1} F^{\lambda\lambda}_{(k,k)} = N\frac{e^2/m^2}
{(k^0)^2} \left[\frac{2}{3} \sigma^2 + \delta^2 \cdot sin^2 \theta
\right]
\label{eq:R}
\ee
where $\ \theta\ $ is the angle between the photon and the beam
directions and the polarization sum is calculated using the well-known
identity:
\be
\sum^2_{\lambda=1} \bepsilon^i_{\lambda}(k) \bepsilon^j_{\lambda}(k)
= \delta^{ij} - {\bf n}^i{\bf n}^j \
\label{eq:sp}
\ee
with $\ {\bf n}={\bf k}/|{\bf k}|\ $.

To calculate the double inclusive cross-section (as well as the higher
order inclusive spectra) one has to know higher order current
correlators. In our case when all proton-neutron collisions are
considered to be independent from each other and the number of the
participating protons is sufficiently large $\ N > 10\ $, we can apply
the central limit theorem and express the higher order current
correlators through the first and second ones. Assuming $\
<J^{\lambda}(k)>=0\ $ (no coherence) the double inclusive
cross-section is represented through the sum of products of the
two-current correlators
\ba
&&\rho_2(k_1,k_2) \ = \ \sum^2_{\lambda_1,\lambda_2=1}
<J^{\lambda_1}(k_1) J^{\lambda_1}(-k_1) J^{\lambda_2}(k_2)
J^{\lambda_2}(-k_2)> =
\nonumber \\
 &&\sum^2_{\lambda_1,\lambda_2}
\left\{ F^{\lambda_1 \lambda_1}(k_1,k_1)
F^{\lambda_2 \lambda_2}(k_2,k_2) + F^{\lambda_1 \lambda_2}(k_1,k_2)
F^{\lambda_2 \lambda_1}(k_2,k_1) + F^{\lambda_1 \lambda_2}(k_1,-k_2)
F^{\lambda_2 \lambda_1}(-k_2,k_1) \right\}
\nonumber \\ &&
=\rho_1(k_1) \rho_1(k_2) +
\left[ \sum^2_{\lambda_1,\lambda_2=1} F^{\lambda_1 \lambda_2}(k_1,k_2)
F^{\lambda_2 \lambda_1}(k_2,k_1) + (k_2 \leftrightarrow - k_2) \right]
\label{eq:rho2}
\ea
and the polarization sum
\ba
\sum^2_{\lambda_1 \lambda_2=1} F^{\lambda_1\lambda_2}(k_1,k_2)
F^{\lambda_2\lambda_1}(k_2,k_1) = |\tilde{f} (k_1-k_2)|^2 \cdot N^2
\frac{e^4/m^4}{(k^0_1 k^0_2)^2} \times \qquad \qquad
\qquad \label{eq:sum} \\
\left\{ \frac{\sigma^2}{9}(1+\cos^2 \psi)
+ \delta^2
\sin^2\theta_1 \sin^2 \theta_2 +
 \frac{2}{3} \sigma \delta [1 - cos^2 \theta_1 - \cos^2\theta_2 + \cos
\psi
\cos \theta_1 \cos \theta_2] \right\} \nonumber
\ea
is performed with the help of (\ref{eq:F3}), (\ref{eq:sp}).

The general expression for the second order correlation function is
defined by
\be      \label{C2:def}
C_2(k_1,k_2)\ =\ \frac{\rho_2(k_1,k_2)}{\rho_1(k_1)\rho_1(k_2)}
\ee
and has a complicated angular dependence, but the two limiting cases
1) $\sigma \gg \delta$ and 2) $\sigma \ll \delta$ lead to very simple
expressions:

1. For the case $\sigma \gg \delta$ we have:
\be
C_2(k_1,k_2|\sigma\neq0,\delta=0)\ =\ 1 + \frac{1}{4} (1 + \cos^2
\psi)\left[|\tilde{f}(k_1-k_2)|^2 + |\tilde{f}(k_1+k_2)|^2\right]
\label{eq:c}
\ee
which is the result derived in \cite{N} and which gives for the
intercept
\be
C_2(k,k) = \frac{3}{2} + \frac{1}{2}|\tilde{f}(2k)|^2 \ .
\label{eq:c2}
\ee

2. The opposite case $\sigma \ll \delta$ leads to another formula:
\be
C_2(k_1,k_2|\sigma=0,\delta\neq0) = 1 + |\tilde{f}(k_1-k_2)|^2 +
|\tilde{f}(k_1+k_2)|^2 \ ;
\label{eq:c3}
\ee
with the intercept exceeding 2:
\be
C_2(k,k) = 2 + |\tilde{f}(2k)|^2 \ .
\label{eq:c4}
\ee
For hard photons $\ |\tilde{f}(2k)|^2\ll 1\ $ (cf. \cite{N}) and one
can neglect this contribution to the intercept while for soft photons
$\tilde{f}(2k)$ is non-neglegible and in the limit $k=0,\
\tilde{f}(0)=1$ so that $C_2(k,k)=3$. The real situation (when both
$\sigma$ and $\delta$ contribute) is between (\ref{eq:c}) and
(\ref{eq:c3}) and exhibits a more complicated angular behaviour than
the considered above limiting cases
\cite{RPW}.  For instance the general expression for the intercept which
follows directly from (\ref{eq:R}), (\ref{eq:rho2}), (\ref{eq:sum})
\be
C_2(k,k)= \frac{\int d\Omega \rho_2(k,k)}{\int d\Omega \rho_1^2(k)} =\
1+\frac{1}{2}[1+|\tilde{f}(2k)|^2]\cdot \left[ 1+
\frac{1.2\delta^2}{\sigma(\sigma+2\delta)+1.2\delta^2}  \right]
\label{max:gen}
\ee
depends on both parameters $\sigma$ and $\delta$ and varies in the
range between $3/2$ and $3$. The solid angle integration over all
possible orientations of the photon momentum ${\bf k}$ corresponds to
a $4\pi$ detector.

The function $f(x)$ and its Fourier transform $\tilde{f}(k)$) reflects
the space-time properties of the photon source. It depends in
principle on three constants: the time duration $R_0$, the
longitudinal radius $R_l$ and the transverse radius $R_t$. One can
propose a concrete parametrization for the source geometry
$\tilde{f}(k)$ and then fit the inclusive data using (\ref{C2:def}).
However, because of insufficient statistics one has usually to limit
oneself to a smaller number of parameters.
In particular the choice of two parameters $T=R_0$ and $R=R_l=R_t$ is
good enough for the present state of the art.  With this assumption we
can perform analytically the angular integration over $\theta_1$ and
$\theta_2$ in (\ref{eq:R}) and (\ref{eq:rho2}), (\ref{eq:sum}) keeping
constant the angle between the two photons ${\bf n}_1 \cdot {\bf n}_2
=cos\psi=const$. The corresponding integration
\be  \label{dmu}
\int d\mu \ =\ 2\pi\int\limits^{\pi}_0 sin\theta_1 d\theta_1
\int\limits^{2\pi}_0 d\varphi
\ee
extends over the rotations of ${\bf n}_2$  around  ${\bf n}_1$
$(d\varphi)$ and over all orientations of ${\bf n}_1$ around
the beam direction. After some algebra we are left with the general
expression for
the  Bose-Einstein correlation function
\ba
&& C_2(k_1,k_2)\ \equiv \ \frac{\int d\mu \rho_2(k_1,k_2)}{\int d\mu
\rho_1(k_1)\rho_1(k_2)}\ =\ 1 \ + \label{C:int}  \\
&&\frac{1}{4}\left\{\left( |\tilde{f}(k_1-k_2)|^2 + |\tilde{f}(k_1+k_2)|^2
\right)\cdot \left[ 1+cos^2\psi
+\frac{0.3\delta^2(3+cos^2\psi)(3-cos^2\psi)}{\sigma(\sigma+2\delta)
+0.3\delta^2(3+cos^2\psi)}\right]  \right\}
\nonumber
\ea
which has a pronounced dependence not only on the space-time
characteristics $R_0,\ R$ but also on the angle $\psi$ and the
``dynamical'' constants $\sigma,\ \delta$. The limiting cases
(\ref{eq:c}) and (\ref{eq:c3}) as well the intercept formula
(\ref{max:gen}) can be rederived directly from (\ref{C:int}).
\section{Partially coherent sources}

In general, photon sources are not totally chaotic but may contain
also a coherent component $<J^{\lambda}(k)>\equiv I^{\lambda}(k)\ne
0$\ so that the total current
$J^{\lambda}_{T}=J^{\lambda}_{ch}+I^{\lambda}$ leads to a partially
coherent field. There are several mechanisms which are responsible for
coherence: collective deacceleration of the initial nuclei, collective
flow, coherent radiation from nuclear fragments, etc. The possibility
to investigate such collective phenomena is an interesting and
important part of heavy-ion physics.  In this chapter we discuss
phenomenologically the influence of the coherent part of the electric
current on the photon correlation function.  The single and double
inclusive cross-sections (see (\ref{eq:R}) and (\ref{eq:rho2}) ) read:
\ba
&&\rho_1(k)\ =\  \sum^2_{\lambda=1}\left( F^{\lambda \lambda}(k,k)\
+\ |I^{\lambda}(k)|^2 \right) \quad , \label{rho1:coh}   \\
&&\rho_2(k_1,k_2)\ =\ \sum^2_{\lambda_1,\lambda_2=1}<J^{\lambda_1}(k_1)
J^{\lambda_1}(-k_1)J^{\lambda_2}(k_2)J^{\lambda_2}(-k_2)> =
\rho_1(k_1)\rho_1(k_2)+ \label{rho2:coh}   \\
&&+ \sum^2_{\lambda_1,\lambda_2=1} \left\{  |F^{\lambda_1 \lambda_2}
(k_1,k_2)|^2 + 2{\cal R}{\rm e}[I^{^{*}\lambda_1}(k_1)F^{\lambda_1\lambda_2}
(k_1,k_2)I^{\lambda_2}(k_2)] + (k_2 \lraw -k_2)
\right\} \quad . \nonumber
\ea
As before we assume axial symmetry around the beam direction and
parametrise $I^{\lambda}(k)$ as follows:
\be
I^{\lambda}(k)\ =\ \frac{ie}{m k^0} \sqrt{N} S(k_0)\bell \cdot
\bepsilon^{\lambda}(k)\quad .
\label{I}
\ee
Using  (\ref{eq:F3}),  (\ref{eq:sp}) and  (\ref{I}) we get
\ba
&& \sum^2_{\lambda=1}|I^{\lambda}(k)|^2\ =\ N
\left( \frac{e/m}{k^0} \right)^2 |S(k_0)|^2 sin^2\theta \quad ,
\label{II} \\
&& \sum^2_{\lambda_1,\lambda_2=1}I^{^{*}\lambda_1}(k_1)F^{\lambda_1
\lambda_2} (k_1,k_2)I^{\lambda_2}(k_2)= \tilde{f}(k_1-k_2)N^2 \left(
\frac{e^2/m^2}{k^0_1 k^0_2} \right)^2 S^{^{*}}(k^0_1)S(k^0_2) \times
\nonumber \\
&& \left[ \frac{\sigma}{3}(1-cos^2\theta_1-cos^2\theta_2+cos\psi
cos\theta_1cos\theta_1) + \delta sin^2\theta_1sin^2\theta_2 \right]
\label{IFI}
\ea
which together with $F^{\lambda \lambda}(k,k)$ and $F^{\lambda_1
\lambda_2}(k_1,k_2)F^{\lambda_2 \lambda_1}(k_2,k_1)$ (see
(\ref{eq:R}),(\ref{eq:sum})) determine completely the single and
double inclusive spectra (\ref{rho1:coh},\ref{rho2:coh}).  As compared
to the completely chaotic case there is now one more function $S(k_0)$
entering in $\rho_1(k)$ and $\rho_2(k_1,k_2)$.

Significant simplifications can be achieved again assuming
$R_l=R_t=R$.  The time dependence of the photon source is still
arbitrary. It is then possible again to perform explicitly the
integrations over all the angles except for $\psi$ and write the
photon-photon correlation function in the compact form
\ba
&&C_2(k_1,k_2)\ \equiv \ \frac{\int d\mu \rho_2(k_1,k_2)}{\int d\mu
\rho_1(k_1)
\rho_1(k_2)} \ =  \label{C2:coh} \\
&&=\ 1\ +\ \left\{ \lambda^2(k_0)A|\tilde{f}(k_1-k_2)|^2 +
2\lambda(k_0)(1-\lambda(k_0))B{\cal R}{\rm e}\tilde{f}(k_1-k_2) + (k_2
\leftrightarrow -k_2) \right\} \nonumber
\ea
where we have introduced a new set of parameters:
\ba
\lambda(k_0)&=& \frac{\sigma + \delta}{\sigma + \delta + |S(k_0)|^2}\ ,
\qquad x= \frac{\sigma}{\sigma + \delta}\ , \label{par:coh} \\
A&=& \frac{1}{4}\cdot \frac{1+cos^2\psi+(1-x)^2(13+cos^2\psi)/5}%
{1+(1-x\lambda)^2(3cos^2\psi-1)/10}\ , \nonumber \\ B&=&
\frac{1}{4}\cdot \frac{1+cos^2\psi+(1-x)(13+cos^2\psi)/5}%
{1+(1-x\lambda)^2(3cos^2\psi-1)/10}\ . \nonumber
\ea
Here $\lambda(k_0)$ is of the chaoticity parameter which is the ratio
of the number of chaotically produced photons with energy $k_0$ and
their total number $\lambda(k_0)=<n(k_0)>_{ch}/<n(k_0)>~.$ We assume
also that on the scale important for the Bose-Einstein study
$(|\bold{k}_1 -\bold{k}_2|\sim 1/R)$ the function $\lambda(k_0)$ does
not vary significantly: $|(k^0_1-k^0_2)d\lambda(k^0)/dk^0| \ll
(\lambda(k^0_1)+\lambda(k^0_2))/2~.$ The other parameter which
influences strongly the behaviour of the correlation function
$C_2(k_1,k_2)$ is the isotropy parameter $x$ (see (\ref{par:coh})).
Both these parameters $\lambda$ and $x$ can be extracted from the
analysis of the angular distribution of the radiation and the value of
the intercept $C_2(k,k)$:
\ba
&& \rho_1(\bold{k})/<n(k_0)>\ =\ \frac{1}{4\pi}[x\lambda(k_0)+
\frac{3}{2}(1-x\lambda(k_0))sin^2\theta] \quad , \label{ang} \\
&&C_2(k,k)=1+\frac{(1+|\tilde{f}(2k)|^2)/2}{5+(1-x\lambda)^2}\left\{
\lambda^2(k_0)[5+7(1-x)^2]+2\lambda(1-\lambda)[5+7(1-x)]\right\}
\  \quad \label{xl:coh}
\ea
(for the case of hard photons $|\tilde{f}(2k)|^2 \ll 1$ and one can
neglect this contribution in (\ref{xl:coh})).  After this has been
done the only unknown function in (\ref{C2:coh}) is the space-time
distribution function of the source $f(x)$ (or $\tilde{f}(k)$) which
can now be obtained by fitting the experimental data. This function
has the physical meaning of the space-time distribution of the
radiating region and reflects the geometry of the early stages of the
collision.

We would like to stress once again that both the parameters $\lambda$
and $x$ strongly influence $C_2(k_1,k_2)$ and the knowledge of only
the one of them (e.g. $\lambda$) is not enough to determine in a
unique way the two-photon correlation function. For instance, one can
check that all the $\lambda, \ x$ pairs, for which
$\lambda=(\sqrt{2}-1)/(\sqrt{2}-x)$ and $0\le x\le 1~,$ lead to the
same intercept $C_2(k,k)=3/2~.$


\section{Exponential fall of the bremsstrahlung amplitude}
\label{ef}
The experimental observations  \cite{GSI,C} show that the one-particle
inclusive spectrum of brems\-strahlung photons has exponential form
suggesting that the underlying proton current reads:
\be
j^{\lambda}(k) = \frac{ie}{mk^0}{\bf p}\cdot\bepsilon_{\lambda}(k)
\cdot {\rm exp}[-k_0/(2E_0)]
\label{A:j}
\ee
rather than (\ref{eq:j}).  The single and double inclusive
cross-sections calculated with (\ref{A:j}) instead of (\ref{eq:j}) can
be obtained from the previous results multiplying them by the
corresponding powers of ${\rm exp}[-|k^0_1|/(2E_0)]$ and ${\rm
exp}[-|k^0_2|/(2E_0)]~.$ \\ The homogeneous functions like the
two-photon correlation function (\ref{C2:def}) and the angular
distribution of radiation (\ref{ang}) remain unchanged and all the
conclusions about the Bose-Einstein correlations obtained above hold.

In the following we shall derive the non-relativistic classical
trajectory of a proton which leads to the mentioned above current
(\ref{A:j}). \\ The current in momentum space is defined through the
proton trajectory as follows:
\be
j^\lambda(k)=\epsilon^{\lambda}_\mu(k)\tilde{j}^\mu(k)= -e
\int\limits^{+ \infty}_{-\infty}dt\,\,
{\rm exp}\{i[k_0t-\bold{k}\bold{r}(t)]\}
\left(\bepsilon_{\lambda}\cdot \bold{v}(t)\right) \quad .
\label{A:j2}
\ee
In non-relativistic case $|\bold{v}(t)|\ll 1$ and as a consequence
$k_0t-\bold{k}\bold{r}(t)\cong k_0t~. $ Therefore $j^{\lambda}(k)$
reduces to the time Fourier transform of the velocity.
Using the identity
\be
\int\limits^{+ \infty}_{- \infty}dt{\rm e}^{i\omega t}
\frac{1}{2}\left[ 1 -\frac{2}{\pi}arctg(2E_0t) \right]=
\frac{-i}{\omega -i\epsilon}{\rm exp}[-|\omega|/(2E_0)]
\label{A:id}
\ee
one finds the trajectory
\be
\bold{v}(t)\ =\ \frac{\bold{v}_0}{2}\left[1\ -\
\frac{2}{\pi}arctg(2E_0t)\right]
\label{A:v}
\ee
leading to the proton current (\ref{A:j}). The standard formula
(\ref{eq:j}) corresponds to the special case $E_0 \rightarrow +
\infty$ when the proton trajectory is described by the step function
$\bold{v}(t)=\bold{v}_0\Theta(-t)~$.  The finite value of $E_0$
reflects more smooth then step-like deacceleration of the proton with
the characteristic stopping time $\tau \sim 1/(2E_0)$ and stopping
length $l \sim v_0 \tau$. For instance, for the projectile energy $45
\ MeV/u$ (slop-parameter $E_0=18 \ MeV$ \cite{C}) one gets $\tau=5.6 \
fm/c$ and $l=1.7 \ fm$ which seem to be quite reasonable.

\section{Summary}

In this paper the production of photons is analysed in the framework
of quasi-classical approximation. Our consideration is valid in the
region of low and intermediate colliding energies up to $1000 \
MeV/u$. We assume that the hard photons $(E_{\gamma}\ge 25 MeV)$ are
produced in independent proton-neutron collisions\footnote{the photon
emission from the proton-proton collisions has the quadrupole nature
and therefore is highly suppressed with respect to dipole radiation
from the proton-neutron collisions.} (see Chapter 2) and the whole
system has the axial symmetry with respect to the beam direction. Then
we derive the expressions for the single and double inclusive
cross-sections and for the two-photon correlation function as well
(see Chapters 3,4). We show that the behaviour of the photon-photon
correlation function depends not only on the space-time properties of
the collision region (function $\tilde{f}(k_1-k_2))$ but also on
dynamics of the proton-neutron scattering in matter (parameters
$\sigma,\ \delta$ and the amount of chaoticity $\lambda$).  So far the
photon intensity interferometry can be considered as an indirect way
to check the dynamical properties of the heavy-ion system as well as
to obtain the space-time information about a collision.

It turns out that the maximum value of the correlation function (the
intercept $C_2(k,k)$) is very sensitive to the details of the
proton-neutron scattering and varies generally speaking in the
interval $\ 1 \le C_2(k,k) \le 3\ $ (see (\ref{xl:coh})). If we deal
with so hard photons that $|\tilde{f}(2k)|^2 \ll 1\ $ the intercept
finds itself in more narrow range $\ 1 \le C_2(k,k) \le 2\ $ (which is
nevertheless wider then the one obtained in \cite{N} $\ 1 \le C_2(k,k)
\le 1.5\ $). The reason why the intercept can exceed the value 1.5\
(see \cite{N}) is strongly connected with the beam-direction
anisotropy specific for heavy-ion collisions. \\ The photons produced
with the close momenta can have the correlations in their
polarisations. As far as the photons with the same polarisation obey
the Bose-Einstein effect like the scalar bosons (e.x. $\pi^0$) and
ones with the perpendicular polarisations behave like nonidentical
particles, these correlations in polarisations affect positivly on the
Bose-Einstein peak increasing intercept. The presence of coherence
leads to decreasing of the correlation effect. In order to study the
two-photon correlation function including the influence of the
anisotropy and the coherent contribution as well (\ref{C2:coh}) one
needs additional information which can be obtained from the angular
distribution of the radiation (\ref{ang}) and the intercept value
(\ref{xl:coh}) analysed together in order to extract the values
$\lambda$ and $x$.  After it has been done the only unknown quantity
in (\ref{C2:coh}) is the Fourier transform $\tilde{f}(k)$ of the
space-time probability distribution $f(x)$ of the photon source which
reflects the geometry of the early stages of the heavy-ion collision.

We would like to acknowledge fruitful comments by Y.Schutz,
I.V.Andreev, G.R\"opke, T.Alm, J.Clark.\\ This work was supported in
part by the Gesellschaft f\"ur Schwerionenforschung, Darmstadt.

\end{document}